\documentclass[12pt]{article}
\usepackage{amsmath,amssymb}
\usepackage{graphics,epsfig}
\usepackage[dvips]{color}
\usepackage[letterpaper,colorlinks,breaklinks,bookmarks]{hyperref}
\begin{document}
\baselineskip 18pt
\title{Enhanced four-wave mixing via
elimination of inhomogeneous broadening by coherent driving of
quantum transition with control fields}
\author{Alexander K. Popov$^{\star}$ and Alexander S. Bayev\\
{\em Institute of Physics of Russian Academy of Sciences and}\\
{\em Krasnoyarsk State University, 660036 Krasnoyarsk, Russia.}}
\date{}
\maketitle
\begin{abstract}
\baselineskip 18pt
We show that atoms from wide velocity interval
can be concurrently involved in Doppler-free two-photon resonant
far from frequency degenerate four-wave mixing with the aid of
auxiliary electromagnetic field.  This gives rise to substantial
enhancement of the output radiation generated in optically thick
medium. Numerical illustrations addressed to typical experimental
conditions are given.
\end{abstract}

{PACS number(s):  42.50.Gy, 42.65.Dr,  42.65.Ky}
\vspace{5mm}

Strong optical resonances inherent to free atoms and molecules
are negated by the fact that only small fraction of the species
can be concurrently resonance coupled in warm bulk gases. This
is because of Maxwell distribution of the Doppler shifts of
their resonances. Doppler-free (DF) coupling is usually
achievable under equal frequency counter-propagating weak waves
in two-photon-resonant ladder schemes. Because of the phase
matching requirements, such schematic can not be implemented for
far from frequency degenerated four-wave mixing (FWM). Second,
the conditions of intermediate one-photon quasi-resonance can be
very seldom satisfied in this case. Third, DF coupling vanishes
with the growth of not only frequency difference but with
strengths of the coupled fields too due to ac-Stark effect.
Forth,  DF coupling can not be routinely achieved in Raman
schemes, like considered in this paper, because of the inherent
frequency difference. Fifth, DF absorption does not indicate
readily achievable DF FWM polarization in general case.
Moreover, it is not obvious that increased FWM polarization
would result in enhanced output of generated radiation, because
of accompanying increased absorption.

This paper is aimed at demonstrating that the outlined
limitations in optical physics can be removed by making use
quantum coherence processes induced by an auxiliary intense
control field. Considered effects lead to concurrent
contribution of atoms from a wide velocity interval to the
induced resonance and to eliminating it's Doppler broadening
under moderate light intensities. We investigate two-
photon-resonant Raman-type FWM, controlled by an auxiliary
driving field. The explicit formulae for power- dependent
absorption/gain indices and for nonlinear FWM susceptibilities,
accounting for interplay of power and Doppler shifts of the
resonances and illustrating the major idea of the proposed
method are derived. In order to avoid accompanying population
transfer, which would complicate the formulae and would mask the
major effect under consideration, the field coupled to the
ground state is assumed to be weak.

A possible achievement  of sub-Doppler resolution using intense
control field was shown in \cite{Feok,Coh,Tal,Agarw,Baev}.  We
propose a novel scheme, which enables to control FWM coupling
with the aid of auxiliary electromagnetic (EM) field, taking no
part in the FWM process itself.  Accompanying increase of
absorption of the fundamental radiation is considered too. As
the outcome, substantial enhancement in quantum conversion
efficiency in optically thick Doppler-broadened medium is shown.
Numerical illustrations are given for the model, relevant to the
FWM experiments with sodium dimer vapors \cite{Wlg}. As a matter
of fact that detuning from the intermediate resonance are larger
than the Doppler width of the transition and the populated
ground level is coupled to weak fields only, none of the CPT or
EIT effects, usually employed for the enhancement of resonant FWM
\cite {1}, are involved in the proposed technique.
\begin{figure}[!t]
\begin{center}
\includegraphics[width=0.32\textwidth]{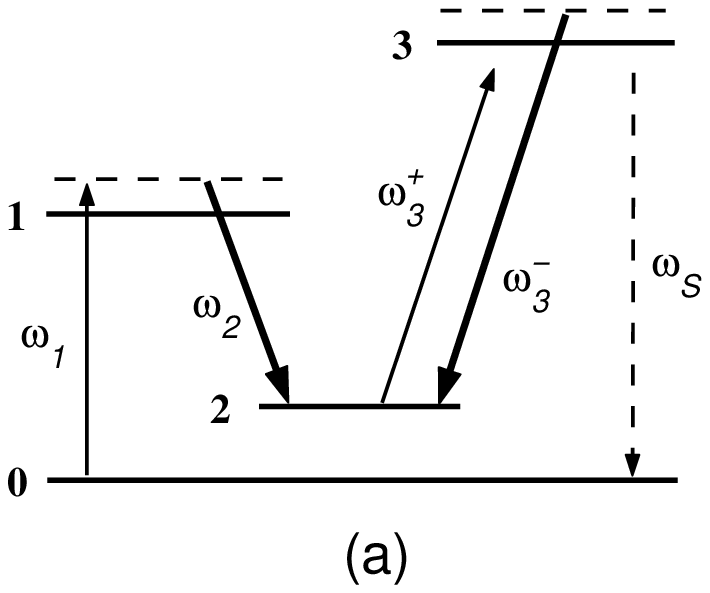}
\includegraphics[width=0.32\textwidth]{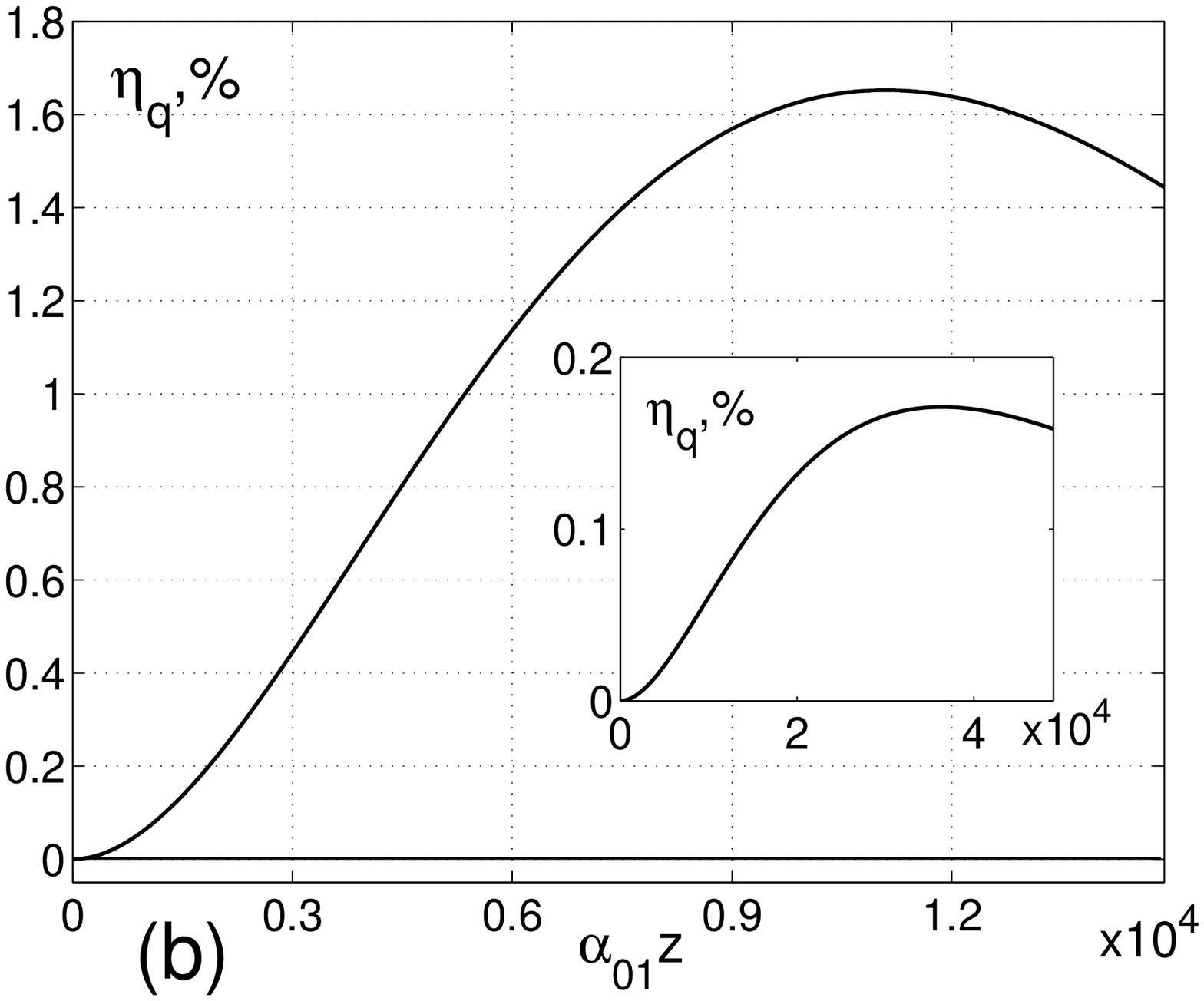}
\includegraphics[width=0.32\textwidth]{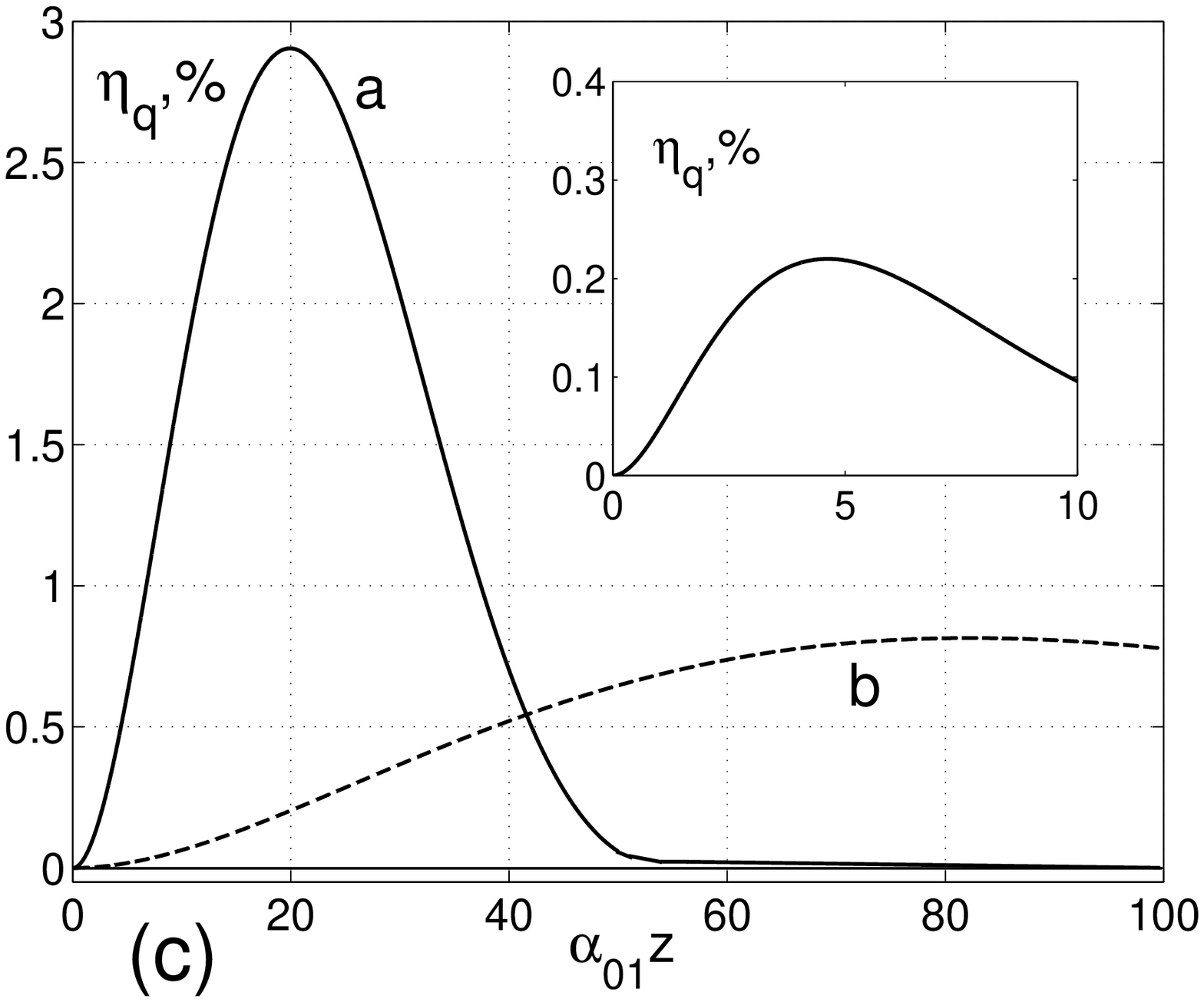}
\end{center}
\caption{\label{pro} Coupling scheme ({\bf (a)}) and enhancement of quantum conversion
efficiency in optically thick medium with the aid of counter-propagating control
field ({\bf (b)} and {\bf (c)} ). ($\alpha_{01}z$ -- optical thickness of the medium
at frequency $\omega_{01}$ under $E_2=E_3^{-}(\omega_3^{-})=0$).   {\bf (b)}:
$G_{12}=128.5$ MHz, $ G_{23}^{-}=25.2$ GHz, $\Omega_2=92.3$ GHz, $\Omega_3^{+}=-7.3$
GHz, $\Omega_3^{-}=73.2$ GHz; inset:  same but $E_3^{-}=0$.  {\bf (c)}: $G_{12}=74.2$
MHz, $ G_{23}^{-}=635.8$ MHz, $\Omega_2=2.3$ GHz, $\Omega_3^{+}=-1.96$ GHz,
$\Omega_3^{-}=1.83$ GHz; plot $b$: same but $E_3^{-}=0$; inset:  $G_{12}=742.2$ MHz, $
G_{23}^{-}=0$, $\Omega_2=2.3$ GHz, $\Omega_3^{+}=0$. In all cases $ G_{23}^{+}=5.78$
MHz, $\omega_1$ is set to induced Doppler-free resonance (for insets -- to the
maximum of nonlinear susceptibility).}
\end{figure}

Consider FWM process $\omega_1-\omega_2+\omega_3^{+}=\omega_S\equiv\omega_4$ and
transition configuration (fig.1a), similar to those studied in the experiments
\cite{Wlg}.  However, in our case the EM radiation ${E}_{3}(t,{z})$ consists of two
components:  weak ${ E}_{3}^{+}(\omega_3^{+})$ and  counter propagating strong one ${
E}_{3}^{-}(\omega_3^{-})$. Their frequencies can be the same or different. Strong
radiation  $E_2$ and weak $E_1$ co-propagate in the same direction as $E_3^{+}$. Only
lower level remains populated, because $E_1$ is assumed so weak, that the populations
can not be driven.  Density matrix equations in the interaction representation are:
\begin{eqnarray}
&L_{01}\rho_{01}=i\{\rho_{00}V_{01}+\rho_{02}V_{21}\},&\label{rho1}\\
&L_{03}\rho_{03}=i\{\rho_{00}V_{03}+\rho_{02}(V_{23}^{+}+V_{23}^{-})\},
&\label{rho2}\\
&L_{02}\rho_{02}=i\{\rho_{01}V_{12}+\rho_{03}(V_{32}^{+}+V_{32}^{-})\},
&\label{rho}
\end{eqnarray}
where $\displaystyle L_{ij}={\partial}/{\partial t}+{\bf v}{\bf\nabla}+
\Gamma_{ij}$, $\displaystyle V_{ij}=G_{ij}\exp\{i(\Omega_{i}t- k_{i}z) \}$,
$V_{23}^{\pm}=G_{23}^{\pm}\exp\{i(\Omega_{3}^{\pm}t\mp k_{3}^{\pm}{z}) \}$,
$\displaystyle G_{ij}=-{E_j d_{ij}}/{2\hbar}$, $\displaystyle
G_{23}^{\pm}=-{E_3^{\pm} d_{23}}/{2\hbar}$ - are coupling Rabi frequencies,
$\displaystyle \Omega_{i}$ -  are corresponding resonance detunings (e.g.,
$\Omega_1=\omega_1-\omega_{01}$), $\displaystyle \Gamma_{ij}$ - homogeneous
half-widths of the transitions.

As follows from (\ref{rho1}) -- (\ref{rho2}), induced atomic coherence
$\rho_{02}$ gives rise to the components in polarizations,  responsible for
novel effects  in absorption and generation of the radiations under
consideration. In the lowest order on the strength of the weak fields
solution of the equations (\ref{rho1}) -- (\ref{rho}) can be found in the
form:\\
$\displaystyle \rho_{02}=r_{02}^{(1)}\exp\{i[(\Omega_{1}-\Omega_{2})t-
(k_{1}-k_{2})z]\}+r_{02}^{(4)}\exp\{i[(\Omega_{4}
-\Omega_{3}^{+})t-(k_{4}-k_{3}^{+})z]\}+\tilde r_{02}^{-}\exp\{i[(\Omega_{4}
-\Omega_{3}^{-})t-(k_{S}+k_{3}^{-})z]\},$
$\displaystyle\rho_{03}=r_{03}\exp\{i(\Omega_{4}t-k_{4}z)+
\tilde{r}_{03}\exp\{i(\Omega_{4}t-k_{S}z)\}+
\tilde{r}_{03}^{-}\exp\{i(\Omega_{4}^{-}t-k_{S}^{-}z)\}$,
$\displaystyle \rho_{01}=r_{01}\exp\{i(\Omega_{1}-k_{1}z)\}+
\tilde r_{01}\exp\{i(\Omega_{1}t-\tilde k_{1}z)\}+
\tilde r_{01}^{-}\exp\{i(\Omega_{1}^{-}t-\tilde k_{1}^{-}z)\}, $
where $\Omega_{4}^{-}= \Omega_{1}-\Omega_{2}+ \Omega_{3}^{-},
 \Omega_{1}^{-}= \Omega_{4}- \Omega_{3}^{-} +\Omega_{2},
 k_S = k_1 - k_2 + k_3^{+},
 k_S^{-} = k_1 - k_2 - k_3^{-}, \tilde k_1 = k_4 - k_3^{+}+ k_2,
 \tilde k_1^{-} = k_4 + k_3^{-}+ k_2$.

Equations for the density-matrix amplitudes become algebraic.
With aid of solution for $\tilde{r}_{03}, {r}_{01}$ and
${r}_{03}$ expressions for the susceptibilities, dressed by the
strong fields $E_2$ and $E_3^{-}$, can be routinely obtained and
presented as:
\begin{eqnarray}
&\tilde\chi_4^{(3)}=\displaystyle -i \frac{d_{01}d_{12}d_{23}d_{30}/8\hbar}
{P_{01} \tilde P_{02}(P_{03}^{+}+|G_{23}^{-}|^2/P_{02}^{-})}, &\label{tr03}\\
&\displaystyle \frac{\chi_{1}(\Omega_1)}{\chi_{10}}=
\displaystyle\frac{\Gamma_{01}}{P_{01}}\frac{P_{03}^{-}P_{02}
+|G_{23}^{-}|^2}{P_{03}^{-}\tilde P_{02}}, &\label{a1}\\
&\displaystyle \frac{\chi_{4}(\Omega_4)}{\chi_{30}}={
\displaystyle\frac{\Gamma_{03}}{P_{03}} \frac{P_{01}^{-}P_{02}^{-}+
|G_{12}|^2}{P_{01}^{-}\{P_{02}^{-}+|G_{23}^{-}|^2 /P_{03} +|G_{12}|^2
/P_{01}^{-}\}}}, &\label{as}
\end{eqnarray}
where $\chi_{01}$ and $\chi_{04}$ are corresponding resonant values under
the strong fields being turned off, $\displaystyle
P_{01}=\Gamma_{01}+i(\Omega_{1}- k_1v) ,
P_{01}^{-}=\Gamma_{01}+i(\Omega_{1}^{-}- k_1^{-} v),
P_{03}=\Gamma_{03}+i(\Omega_{4}-k_{4}v),
P_{03}^{-}=\Gamma_{03}+i(\Omega_4^{-} -k_4^{-}v),
P_{02}=\Gamma_{02}+i[\Omega_{1}-\Omega_{2}-(k_1 - k_2)v],
P_{02}^{-}=\Gamma_{02}+i[\Omega_{4}-\Omega_{3}^{-}-(k_4 + k_3^{-})v],
\tilde {P} _ {02}= P_{02}+|G_{12}|^2
 /P_{01}+|G_{23}^{-}|^2/P_{03}^{-}$,
$v$  is  projection of atom velocity  on $z$. Difference between
$k_1$ and $\tilde k_1$ as well as between $k_4$ and $k_S$ is
neglected here.

With account of absorption but neglecting depletion of
fundamental radiations due to FWM conversion, reduced  equation
for $E_4$ can be written as:
\begin{eqnarray}
dE_4(z)/dz={\rm i}2\pi k_4\tilde\chi ^{(3)}_4E_{1}(0)E_{2}^{*}E_{3}
\exp ({\rm -i}\Delta Kz), & \label{volsis}
\end{eqnarray}
where  $\Delta K=K_{4 }-K_{1}+K_{2}^{*}-K_{3}^{+}$, $K_j = k_j-{\rm i}\alpha
_{j}/2$ -- are complex wave numbers, $\alpha _{j}$ -- power-dependent
absorption indices.  Quantum conversion efficiency ($QCE$) of $E_1$ into $E_4$
along the medium $\eta_{{\rm q}}(z)$ is given by  the expression:
$
\eta_{{\rm q}}(z)={(\omega_1}/{\omega_4)}|E_4(z)/E_1(0)|^2
\exp(-\alpha_4z)\label{qu}.
$

From (\ref{volsis}) one obtains:
\begin{eqnarray}
\eta_{\rm q}= \displaystyle \frac{\omega_1}{\omega_4}
\frac {\big|2\pi\chi_4^{(3)} E_2
E_3\big|^2}{|\Delta K|^2}\exp(-\alpha_4 z) \big|exp(-{\rm i}\Delta K z)
-1\big|^2.\label{qus}
\end{eqnarray}
Pre-exponential factor can be expressed over Rabi frequencies, reduced
nonlinear susceptibility and absorption indices considered below, ratios
of the transition widths and ${|d_{03}|^2}/{|d_{01}|^2}$. The last factor
is proportional to the ratio of the spontaneous relaxation rates. Thus
$QCE$ can be found as absolute value, dependent on the optical thickness of
the medium.

The major physics underlying the proposed technique is as
follows. Modulation of the atomic wave-functions by the driving
fields gives rise to the Autler-Townes splitting, which exhibits
itself in our case as resonance shift.  Besides intensities, the
later depends on detunings of the driving fields and
consequently -- on the atomic velocities.  It turns out that
under appropriate intensities the resonances of atoms at
different velocities can be shifted to the approximately same
position. To illustrate that, consider one-photon detunings,
substantially greater than corresponding Doppler HWHM. Then the
resonance power-shift factors in (\ref{tr03}) - (\ref{as}) can be
presented as: $|G|^2/P\approx (1+ikv/p) |G|^2/p$, where $p$ is
corresponding factor $P$ at $v=0$.  This shows possible control
of the resonance Doppler shifts through the power shifts. More
details can be found in \cite{Feok,Baev}. In the same way a
factor in the denominators of (\ref{tr03}), (\ref{a1}),
indicating dressed two-photon resonance, can be presented as:
\begin {eqnarray}
&\tilde P_{02}\approx \tilde \Gamma _{02} + i\tilde \Omega_{02}-
i\big
\{(| G_{12}| ^ 2/ \Omega _ 1 ^ 2) { k} _ 1 +&\nonumber\\
&+\left (1 + \displaystyle \frac {| G_{23}^{-}| ^2} {(\Omega _ 4^{-}) ^ 2}
\right ) (k_ 1 -k_ 2)  -
\displaystyle\frac {| G_{23}^{-}| ^2} {(\Omega _ 4^{-}) ^ 2} k_3^{-}\big
\}{v},&
\label{p02}
\end{eqnarray}
where $\tilde {\Gamma} _{02}$ and $\tilde {\Omega} _ {02}$ give
half-width and position of the induced resonance.  As follows
from (\ref{p02}), under proper choice of detuning, relative
propagation direction and intensity of the control field
$E_3^{-}$, all Doppler shifts can be compensated by the power
shifts in a such  way, that dependence on $v$ vanishes in the
given linear on $v$ approximation. This indicates trapping  of
all atoms, independent of their velocities in DF dressed
two-photon resonance.  It is seen, that as a matter of fact that
$k_2<k_1$, elimination of Doppler broadening is not possible in
the schematic under consideration with the aid of only driving
field $E_2$. However it becomes possible with an auxiliary
counter-propagating control field $E_3^{-}$, which does not
contribute directly in FWM because of phase mismatch. The
equations (\ref{tr03}), (\ref{a1}) show similar behavior of
absorption index and nonlinear susceptibility near induced
resonance.

While approaching closer to the intermediate resonances, required
intensities become lower, but relative contribution of the
neglected terms, proportional to the higher orders on
$k_iv/\Omega_i$, grows. This leads to decrease of the coherently
coupled velocity interval.
\begin{figure}[!t]
\begin{center}
\includegraphics[width=0.98\textwidth]{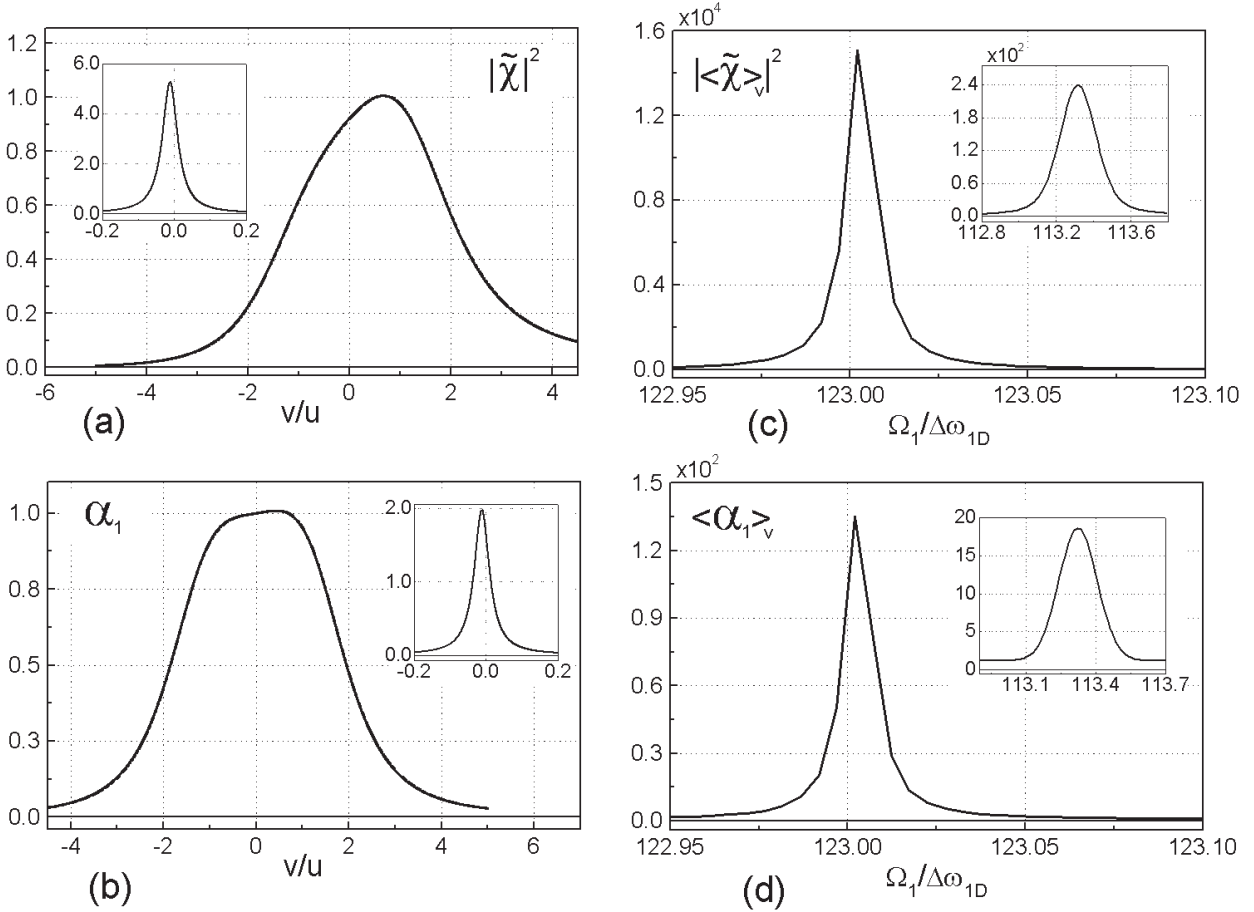}
\end{center}
\caption{\label{chibig}Velocity distribution of the squared modulus of $FWM$ nonlinear
susceptibility ({\bf (a)}) and absorption index at $\omega_1$ ({\bf (c)}) in
Doppler-free resonance; {\bf (b)} and {\bf (d)} -- corresponding velocity-averaged
sub-Doppler resonances (scaled to the corresponding value at the frequency of
control-field induced resonance but under $E_3^{-}(\omega_3^{-})=0$). $u$ - thermal
velocity, $\Delta \omega_{1D}$ -- Doppler HWHM of the transition $01$. The insets --
same functions at $E_3^{-}(\omega_3^{-})=0$. Detunings and intensities are the same
as in fig.  1b.}
\end{figure}

The discussed outcomes can be illustrated with the numerical
model of sodium dimer transitions  \cite{Wlg}:
$\lambda_{01}=661$ nm,  $\lambda_{12}=746$ nm,
$\lambda_{23}=514$ nm and $\lambda_{03}=473$ nm. Corresponding
homogeneous half-widths of the transitions are 20.69, 23.08,
18.30 and 15.92 MHz, Doppler $HWHM$ -- 0.678, 0.601, 0.873 and
0.948 GHz.

Figure 2 depicts contribution  of molecules at different
velocities to the absorption index $\alpha_1(\omega_1) \sim
Re\{\chi_1/\chi_{01}\}$ and to nonlinear susceptibility (trivial
Maxwell envelopes are removed), while conditions of elimination
of Doppler broadening are fulfilled. The figure shows potentials
of coherent coupling of molecules from wide velocity interval
compared with the width of the Maxwell distribution, unlike the
case in the absence of the control field. This gives rise to
strong sub-Doppler resonances. Figure 3 shows modification of the
effects while tuning closer to the intermediate resonances.
Despite the growth of absorption $\alpha_1$, the proposed
manipulating results in substantial increase in output of
generated radiation at $\omega_S$ (fig.1 (b) and (c)).
\begin{figure}[!t]
\begin{center}
\includegraphics[width=0.98\textwidth]{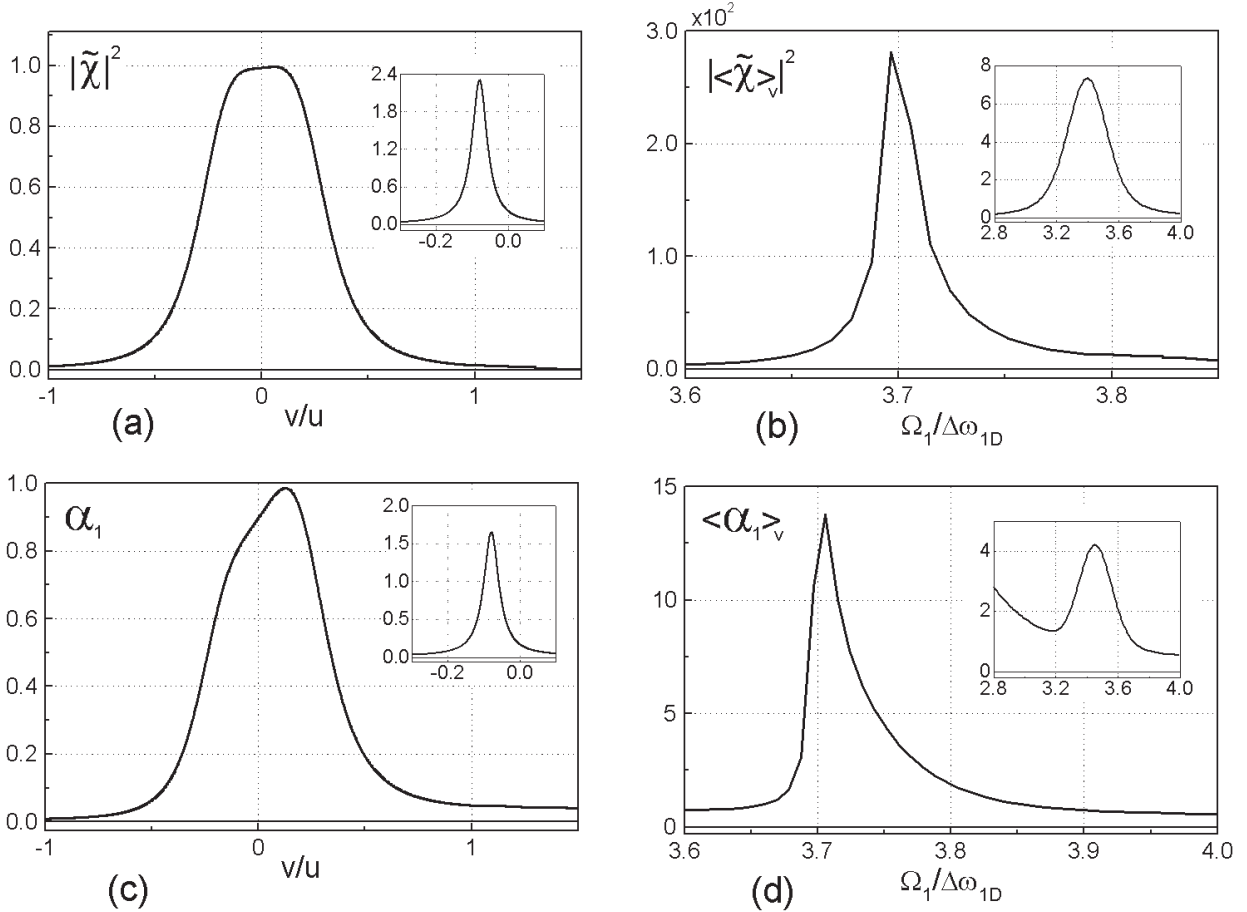}
\end{center}
\caption{\label{chilit}Velocity distribution of the squared modulus of $FWM$ nonlinear
susceptibility ({\bf (a)}) and absorption index at $\omega_1$ ({\bf (c)}) in
Doppler-free resonance; {\bf (b)} and {\bf (d)} -- corresponding velocity-averaged
sub-Doppler resonances (scaled to the corresponding value at the frequency of
control-field induced resonance but under $E_3^{-}(\omega_3^{-})=0$).  The insets:
same profiles at $E_3^{-}(\omega_3^{-})=0$. Detunings and intensities are the same as
in fig.  1c.}
\end{figure}

In conclusion, we show that substantial enhancement in
nonlinear-optical response of a Doppler broadened medium can be
achieved by coherent driving of quantum transitions so that
molecules from wide velocity interval become trapped to one and
the same dressed two-photon Doppler-free resonance. The required
intensities can be decreased by tuning driving frequencies
closer to one-photon resonances, while the coupled velocity
interval decreases too. For ladder-type schemes, where
Doppler-broadening of two-photon resonances is much larger
compared to Raman-like schemes, the considered effects are even
more pronounced.

The authors thank B.Wellegehausen for encouraging discussions.
This work was supported in part by the Krasnoyarsk Regional
Science Foundation, by the Grant 97-5.2-61 in Fundamental
Natural Sciences and by the Russian Foundation for Basic Research
(Grant 99-02-39003).

\end{document}